\documentclass[twocolumn]{aastex701}

\usepackage{times}
\usepackage{amsmath}
\usepackage{multirow}
\usepackage{MnSymbol}
\usepackage{lipsum} 			

\newcommand{\q}[1]{\textcolor{ForestGreen}{\sc #1}}

\shorttitle{3I/ATLAS with SOAR/Goodman HTS}
\shortauthors{Puzia et al.}
\received{July 14, 2025}
\revised{August 4, 2025}
\accepted{August 7, 2025}

\begin{document}

\title{Spectral Characteristics of Interstellar Object 3I/ATLAS from SOAR Observations\footnote{Based on observations collected with the Goodman High Throughput Spectrograph on the 4\,m SOAR Telescope at Cerro Pach\'on Observatory in Chile, under the program allocated by the Chilean Telescope Allocation Committee (CNTAC), program ID CN2025A-96.}}

\author[orcid=0000-0003-0350-7061,gname='Thomas H.', sname=Puzia]{Thomas H. Puzia}
\affiliation{Institute of Astrophysics, Pontificia Universidad Cat\'olica de Chile, Av.~Vicu\~na Mackenna 4860, 7820436 Macul, Santiago, Chile}
\email[show]{tpuzia@astro.puc.cl}  

\author[orcid=0000-0002-5350-0282,gname=Rohan, sname=Rahatgaonkar]{Rohan Rahatgaonkar} 
\affiliation{Institute of Astrophysics, Pontificia Universidad Cat\'olica de Chile, Av.~Vicu\~na Mackenna 4860, 7820436 Macul, Santiago, Chile}
\email{rrohan@uc.cl}

\author[0000-0001-6584-7104,sname=Carvajal,gname='Juan Pablo', sname=Carvajal]{Juan Pablo Carvajal}
\affiliation{Institute of Astrophysics, Pontificia Universidad Cat\'olica de Chile, Av.~Vicu\~na Mackenna 4860, 7820436 Macul, Santiago, Chile}
\email{gjuanpablo@uc.cl}

\author[orcid=0000-0002-4638-1035,gname='Prasanta K.', sname=Nayak]{Prasanta K. Nayak}
\altaffiliation{CATA Post-Doctoral Fellow}
\affiliation{Institute of Astrophysics, Pontificia Universidad Cat\'olica de Chile, Av.~Vicu\~na Mackenna 4860, 7820436 Macul, Santiago, Chile}
\email{pnayak@astro.puc.cl}

\author[orcid=0009-0000-5806-5550,gname=Baltasar, sname=Luco]{Baltasar Luco}
\affiliation{Institute of Astrophysics, Pontificia Universidad Cat\'olica de Chile, Av.~Vicu\~na Mackenna 4860, 7820436 Macul, Santiago, Chile}
\email{baltasarluco@uc.cl}

\begin{abstract}
Interstellar objects (ISOs) provide unique insights into the building blocks and conditions of extrasolar planetary systems.~The newly discovered object, 3I/ATLAS (C/2025 N1), represents the third known ISO after 1I/'Oumuamua and 2I/Borisov. We present initial spectroscopic characterizations of 3I using observations from the {\it Goodman High Throughput Spectrograph} on the 4.1\,m SOAR Telescope in Chile during the night of July 3rd.~The reflectance spectrum of 3I, covering 3700-7000\,\AA\, reveals a red continuum, comparable to extreme trans-Neptunian objects, with a weak UV-optical turnover indicative of complex carbonaceous and irradiated organics.~At the time of observation, when 3I was at a heliocentric distance of $4.4$ AU, we detected no discernible gas emission from canonical cometary species (CN, C$_3$, C$_2$, CO$^+$, [O{\sc i}]). This is in agreement with expectations from our thermal-evolution model, which indicates sublimation-driven activity should commence once 3I/ATLAS approaches smaller heliocentric distances.~Nonetheless, the paradoxical situation of early onset coma without evidence of sublimation tracers, calls for other dust-liberating mechanisms that ancient ISOs may be subjected to at large heliocentric distances.
\end{abstract}


\section{Introduction} 
 \label{ln:intro}
\subsection{1I/'Oumuamua and 2I/Borisov}
Interstellar objects (ISOs) offer a rare transient window into the chemistry of planetary system populations around other stars.~The first ISO, 1I/'Oumuamua (1I), was a $\sim\!100$\,m object with a highly elongated shape inferred from its extreme lightcurve variability and a modestly reddish color, similar to outer Solar System bodies \citep{Jewitt17, Ye17, ISSI19}.~1I showed no detectable coma or outgassing in the optical, yet exhibited a small non-gravitational acceleration likely driven by sublimation of volatiles or other mechanisms \citep{Jewitt17, Micheli18}.~Its reflected spectrum in the visible and near-IR was largely featureless and moderately red, with reported spectral slope $S$ values varying from $\sim\!7$ to 23\%/k\AA\ across different observations \citep{Jewitt17, Meech17, Fitzsimmons18}.~No obvious absorption bands were seen, indicating a lack of strong mineralogic features or ice bands, consistent with an organic-rich irradiated surface analogous to D-type asteroids or cometary crusts \citep{Jewitt23}. 

In contrast, the second ISO, 2I/Borisov (2I), appeared as an active comet \citep{Opitom19,Jewitt19}.~2I developed a visible dust coma and tail, and spectroscopic observations detected typical cometary gas emissions (CN, C$_2$, etc.) in its optical spectrum \citep{Opitom19, Bannister20}. The continuum reflectance of 2I's dust was reddish but not extreme, generally comparable to Solar System comets.~Measured spectral slopes in the optical range spanned roughly $5\!-\!15\%$/k\AA\ in most studies, with some higher values up to $\sim$22$\%$/k\AA\ reported at shorter blue wavelengths \citep{Opitom19, Bolin20}.~Notably, 2I’s composition did show peculiarities in the volatile domain, such as an unusually high CO/H$_2$O ratio, suggesting formation in a cold outer disk environment \citep{Bodewits20, Cordiner20}. However, its reflected continuum remained largely featureless, with no obvious ice absorption features in the $0.5\!-\!2.5$~$\mu$m range, indicating that the grains in its coma were dominated by refractory organics and silicates as in typical cometary dust \citep{Opitom25}.

\subsection{3I/ATLAS}

The new object 3I/ATLAS (C/2025~N1, hereafter 3I) is the third confirmed ISO. It was discovered on July 1st, 2025 at 4.4\,AU heliocentric distance on an inbound trajectory \citep{Denneau25, MPEC2025}\footnote{\url{https://minorplanetcenter.net/mpec/K25/K25N51.html}} by the Asteroid Terrestrial-impact Last Alert System (ATLAS) survey telescope at Rio Hurtado, Chile \citep{Tonry18}. Following its discovery, 3I was subject to intense tracking by observatories worldwide to refine its orbit.~A preliminary orbital solution (using observations spanning May 22–July 6) yielded a perihelion distance $q\!\simeq\!1.358$\,AU and eccentricity $e\!\simeq\!6.15$, firmly establishing the object’s hyperbolic nature \citep{Seligman25, Bolin25, Hopkins25}. These values are even more extreme than those of 1I ($e\!\approx\!1.2$) and 2I ($e\!\approx\!3.37$), making 3I the most hyperbolic object of the three.~Its negative reciprocal semimajor axis translates to a heliocentric $v_{\odot,\infty}\!\simeq\!58$ km/s (compare: $26$ km/s for 1I and 32 km/s for 2I). The inclination $i \approx 175^\circ$ means the comet is on a nearly polar or retrograde orbit relative to the ecliptic plane, in contrast to 2I's $i\approx44^\circ$ (prograde) and 1I’s $i\approx123^\circ$ (retrograde).~Such a high inclination, velocity, and eccentricity indicate that 3I is not gravitationally bound and will escape the Solar System after its perihelion passage on Oct 29.47, 2025, and suggests that it may be the oldest sample so far we have had the opportunity to study from a Milky Way thick-disk star system \citep{Hopkins25}.

Here, we present an early characterization of 3I based on spectroscopic observations with the \textit{Goodman High Throughput Spectrograph} \citep[Goodman HTS;][]{Clemens04} on the 4.1\,m Southern Astrophysical Research (SOAR) Telescope on Cerro Pach\'on in Chile. 

\section{Observations and Data Reduction}  
\label{ln:obs}
The Goodman HTS is a versatile imaging spectrograph optimized for maximum throughput in the UV/optical. Its all--transmissive optical design delivers a circular $7.2\arcmin$ field of view with a plate scale of $0.15\arcsec$/pix.~We used the blue camera, optimized for covering wavelengths from the atmospheric cut-off at $\sim\!3200$\,\AA\ to $8500$\,\AA.~It uses a $4096\!\times\!4096$ pix$^2$ back-illuminated CCD that records photons with high quantum efficiency, yielding a total system throughput of $\sim40\%$ at 5000\,\AA\ when telescope and detector are included. Data was read out with $2\!\times\!2$ binning, a read noise of 4.74\,e$^-$ and 1.4 e$^-$/ADU gain. We used the 1.0\arcsec\ wide long-slit with the 400\,lines/mm volume phase holographic (VPH) grating, providing a dispersion of 1\AA/pix and a spectral resolution of $R\!\simeq\!850$ at 5500\AA.~The weather conditions during the observations were 4\% humidity, 20$^{\circ}$\,C air temperature, 2.8 m/s wind, with a clear sky and 1-1.7\arcsec\ seeing.

\begin{deluxetable}{lcccc}[ht!]
\tablewidth{\textwidth}
\tablecaption{Observing Journal of night 2025-07-03/04 (UT) \label{tab:obs}}
\tablehead{
\colhead{Target} & \colhead{Start (UT)} & \colhead{End (UT)} & \colhead{Airmass} & \colhead{$t_{\rm exp}$[s]}
}
\startdata
   CD-32 9927 & 22:59:17.1 & 23:01:17.1 & 1.04 & 120 \\
   3I/ATLAS & 04:50:33.0 & 05:00:33.0 & 1.05 & 600 \\
   3I/ATLAS & 05:05:44.0 & 05:15:44.0 & 1.06 & 600 \\
   3I/ATLAS & 05:18:41.6 & 05:28:41.7 & 1.08 & 600 \\
   3I/ATLAS & 05:31:16.6 & 05:41:16.6 & 1.10 & 600 \\
   3I/ATLAS & 05:45:00.4 & 05:55:00.4 & 1.13 & 600 \\
   3I/ATLAS & 05:58:15.6 & 06:08:15.5 & 1.16 & 600 \\
   3I/ATLAS & 06:12:13.7 & 06:22:13.6 & 1.20 & 600 \\
   3I/ATLAS & 06:26:01.1 & 06:36:01.0 & 1.24 & 600 \\
   HD168595 & 06:43:33.1 & 06:44:03.1 & 1.23 & 30 \\
\enddata
\tablecomments{During the 3I observations the moon was at 64\% illumination and about 112$^{\rm o}$ away. 3I was moving at a heliocentric $v_\odot\!\simeq\!61$\,km/s and at $v\simeq\!73$\,km/s towards SOAR, resulting in an apparent proper motion of $\mu_{\alpha\cdot\cos{{\rm \delta}}} = -0.0213$\arcsec/sec and $\mu_\delta = 0.0006$\arcsec/sec. 3I was at that time at a heliocentric distance of $r_h\!=$\,4.4\,AU and 3.4\,AU from Earth.}
\end{deluxetable}

For target acquisition we used multiple integrations in imaging mode without the longslit, clearly identifying 3I by its on-sky proper motion (see Tab.~\ref{tab:obs} caption).~After placing the slit mask in the beam, we iteratively centered 3I inside the slit to better than 0.1\arcsec.~With the slit at parallactic angle, we integrated $8\times600$\,sec on 3I with tracking speeds set to the values provided by the JPL Horizons website\footnote{Obtained on UT 2025 July 3 from \url{https://ssd.jpl.nasa.gov/horizons/}.}.~We verified that 3I was centered in the slit using interspersed through-slit images, which never required any correction.~Airmass varied between 1.05 and 1.24 during observations (see Tab.~\ref{tab:obs}).~A solar analog star (HD168595, spectral type G2V) was observed at airmass 1.23 immediately after the 3I sequence to obtain the reflectance spectrum of 3I (see Sect.~\ref{sec:reflectance}).~HgArNe arclamp exposures were taken before and after each sequence, and a flux standard star (CD-32 9927) spectrum was taken at the beginning of the night.

Individual frames were bias subtracted, flat-field corrected, and wavelength calibrated with the Goodman reduction pipeline\footnote{\url{https://github.com/soar-telescope/goodman_pipeline/releases}} \citep{Torres-Robledo20} using calibrations taken within 6 hours of the target observations.~Flux calibration was performed with the flux standard CD-32 9927 observed at the beginning of the night.~After further inspection, we excluded the first and last 3I integration due to excess variance in the continuum as well as blended background stars.~On the remaining calibrated frames, we traced the signal of 3I and the standard stars along the dispersion axis using a sliding-window Moffat function, fitting the light profile center and width along the slit spatial axis as a function of wavelength.~3I and standard-star spectra were extracted using one FWHM apertures along the profile peak trace between $\sim\!3700$\,\AA\ and $7000$\,\AA.~We tested 1 and 2-FWHM extraction apertures and find that the 2-FWHM aperture captures 44\% more total flux than the 1-FWHM aperture (as opposed to the $\sim$33\% that would be expected of a point source), indicating a contribution from an extended component, in agreement with photometric observations that identified the presence of a coma at this early epoch \citep{Seligman25, Chandler25}.~Overall, the two extraction widths yield no significant difference in the spectral slope.~Finally, we obtained 3I's reflectance spectrum using the acquired Solar analog and space-based SOLSPEC spectrum (Sec.~\ref{sec:reflectance} and Fig.~\ref{fig:refspec}).~All spectra including uncertainties are available online\footnote{\url{https://doi.org/10.5281/zenodo.15881487}}.

\section{Analysis} 

\subsection{The Importance of Solar Analog Spectra}
A cautionary note on solar‐irradiance normalization is warranted for 3I’s intensive study during the next months. Solar spectral irradiance is not strictly constant.~Across the 11-yr cycle the bolometric flux rises by only $\sim$0.1\%, yet UV (2000-3000\,\AA) brightening reach 5-10\%, whereas visible–NIR bands vary by $\lesssim\!1$\% \citep{Ermolli13}.~Even these modest, wavelength-dependent shifts matter for high-precision reflectance comparison work (Fig.~\ref{fig:refspec}).

We calibrated our 3I spectrum in two ways: (i) division by the G2V solar twin HD168595 observed at matching airmass, and (ii) division by the SOLSPEC space-based solar spectrum \citep{Meftah18}.~The Solar analog method suppresses most instrumental and atmospheric signatures \citep{Hardorp78,deLeon10}, yet any star–Sun mismatch or untracked solar-cycle drift can imprint spurious structure.~Cross-checking against an absolute solar reference, e.g.~SOLSPEC/ATLAS-3 \citep{Thuillier03}\footnote{We used the SOLSPEC/ATLAS-3 solar spectrum, which was obtained by the third ATmospheric Laboratory for Applications and Science (ATLAS) flown in November 1994 on board the STS-66 Space Shuttle mission, coinciding with low Solar activity.}, helps reveal such biases due to changing Solar activity.~Recent Bouguer–Langley measurements report 6–8\% higher irradiance at 1.6–2.3$\mu$m than SOLSPEC/ATLAS-3, a discrepancy that exceeds expected solar variability and likely reflects calibration differences \citep{Bolsee14}.

Combining contemporaneous Solar analog observations with the latest absolute Solar spectra, therefore, provides the most robust reflectance calibration and error assessment.

\subsection{3I's Reflectance Spectrum Characteristics}\label{sec:reflectance}
\begin{figure*}[ht!]
    \centering
    \includegraphics[width=1\linewidth]{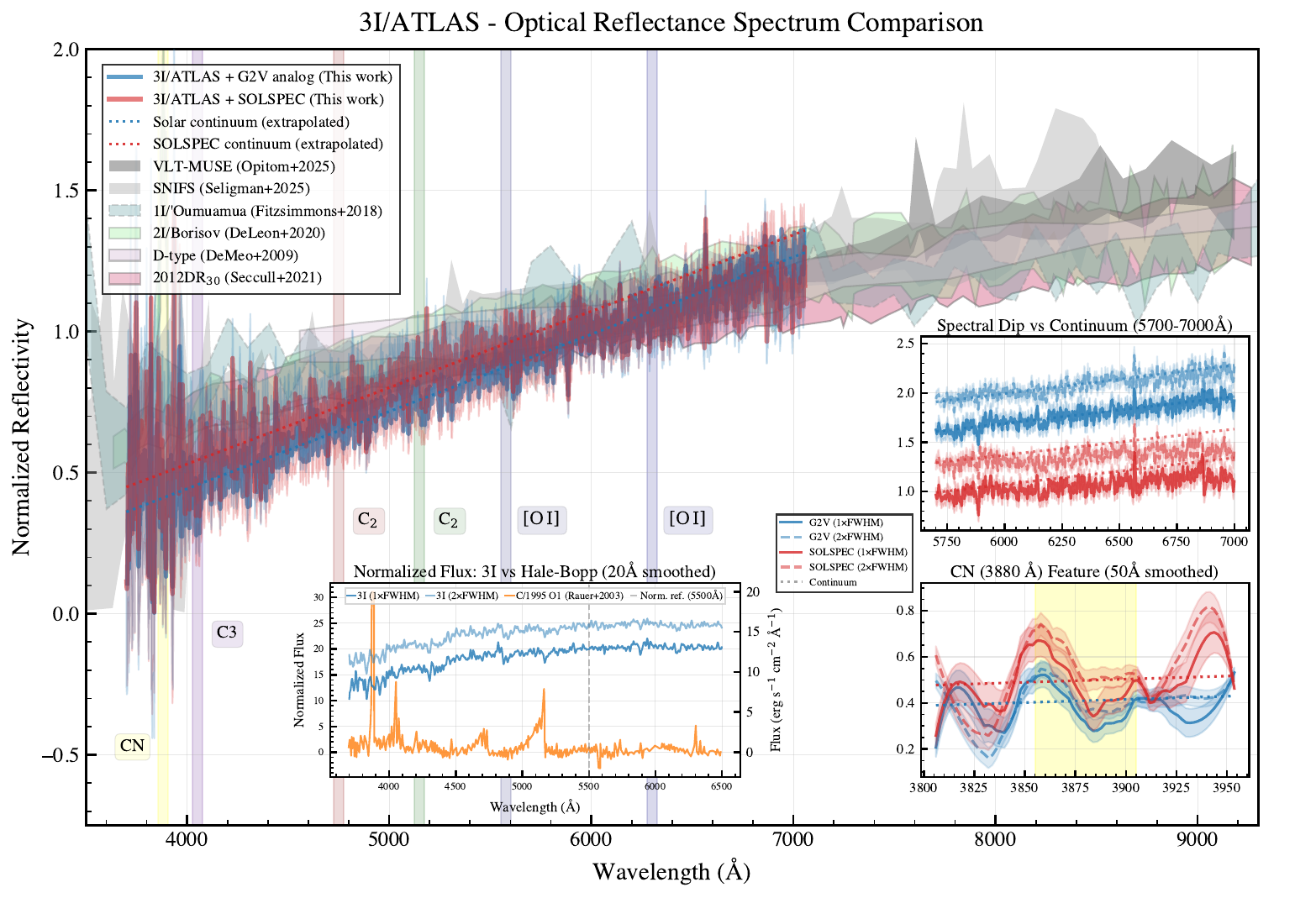}
    \caption{Optical reflectance spectrum of comet 3I obtained at 4.4\,AU compared with literature data for interstellar objects and asteroids. {\it Main panel}: Our 3I spectra obtained using two solar normalization methods are shown: G2V solar analog HD168595 (blue) and SOLSPEC solar spectrum \citep[red;][]{Meftah18}.~Shaded regions show literature reflectance spectra: VLT-MUSE observations of 3I (\citealt{Opitom25}, dark gray), SNIFS observations of 3I (\citealt{Seligman25}, light gray), reflectance of 1I (\citealt{Fitzsimmons18}, cadet/grayish blue), 2I (\citealt{deLeon20}, light green), D-type asteroids (\citealt{DeMeo09}, pink), and 2012DR$_{30}$ (\citealt{Seccull21}, coral).~Colored vertical bands highlight typical molecular emission regions: CN ($3860\!-\!3880$\,\AA, yellow), C$_3$ (4050\,\AA, purple) and C$_2$ (4750\,\AA~and 5150\,\AA, red and green), along with forbidden oxygen transitions \mbox{[O\,\textsc{i}]} (6300\,\AA~and 6364\,\AA, blue). {\it Zoom-in panels}:~({\it Left}) compares the 20\AA-smoothed and normalized (at 5500\,\AA\ and arbitrarily offset) 3I flux spectrum in the 1 and 2-FWHM extraction apertures with the continuum subtracted Hale-Bopp comet emission \citep[right ordinate; from][obtained at $r_h\!=\!3.8$\,AU]{Rauer03}, across the 3500-6500\,\AA~region.~({\it Top right}) shows the 3I reflectance spectral dip region at 6000-7000\,\AA\ with linear continuum fits extrapolated from feature-free region at 4000-5600\,\AA\ to assess absorption features relative to the underlying continuum. Dark and lighter lines are spectra for 1 and 2-FWHM extraction apertures, respectively.~({\it Bottom right}) shows a detailed view of the CN(0-0) vibration band feature region (3780-3940\,\AA) with 50\,\AA\ smoothing applied to reveal the subtle spectral structure. Solid and dashed lines represent the 1 and 2-FWHM extraction apertures, respectively.}
    \label{fig:refspec}
\end{figure*}

There is general agreement between our SOAR/Goodman HTS reflectance spectrum and those acquired with MUSE \citep{Opitom25} and SNIFS \citep{Seligman25} on July 3rd and 4th, respectively.~However, a systematic difference in the red spectral slope is apparent between our SOLSPEC-normalized spectrum (similar to MUSE) and our G2V-normalized spectrum (similar to SNIFS), illustrating how the choice of solar reference spectrum can significantly affect measured reflectance slopes, particularly at wavelengths longer than $\sim\!5500$\,\AA.

Furthermore, while these IFU observations integrate over the entire coma, our \(1\arcsec\) slit samples only a narrow slice through the innermost coma and thus weights flux projected near the nucleus more strongly than farther out.~Although the spectrum remains coma-dominated, such geometric bias, combined with radial gradients in grain size, and, therefore, scattering behavior, can subtly alter the observed continuum slope relative to the IFU data. Consistent with this picture, \citet{Chandler25} model and estimate the ``coma-level'' ($\eta\!=\!2.5\!\pm\!0.5$) on 2025 July 2nd, corresponding to a coma contribution of $\sim\!70\%$ and supporting the expectation that our slit spectrum still predominantly reflects coma flux.

We measure reflectance slopes of ($27.4\!\pm\!1$)\%/k\AA\ at 4000-5500\,\AA\ using our G2V and SOLSPEC solar spectrum, and ($26.7\!\pm\!0.7$)\%/k\AA\ for G2V and ($16.4\!\pm\!0.4$)\%/k\AA\ for the SOLSPEC normalization at 5500-7000\,\AA.~This further demonstrates the importance of acquiring solar analog spectra taken concurrently with the same instrument.

Steep UV slopes flattening into the optical have been observed in D-type asteroids \citep{DeMeo09}, sporadically active centaurs such as Echeclus \citep[at $r_h\!\simeq\!5.8$\,AU;][]{Seccull19}, and 1I \citep{Ye17}. These are attributed to complex organics or radiation damage products that absorb short-wavelength light \citep{Jewitt23}.~Extreme trans-Neptunian objects, such as 2012\,DR$_{30}$ \citep[$r_h\gtrsim14.5$\,AU and $e\!=\!0.99$;][]{Seccull21}, that spend most of their orbital time in interstellar space, show a similar reflectance to 3I's.~The fact that 3I exhibits only a weak UV–optical turnover at 4.4\,AU (see Fig.~\ref{fig:refspec}) would suggest a surface weakly enriched in complex carbonaceous compounds, contrasting its kinship to the most primitive Solar System populations \citep{Rivkin11, Tatsumi20}, but similar to extreme TNOs \citep{Fraser22}.~Even so, we have to keep in mind that ISOs have experienced extremely different thermal and radiation environments compared to the warmer Solar System conditions and its dust coma may be potentially quite different in grain size and composition to the Solar System bodies' surface regolith.~No clear absorption features (e.g.~ice or mineral bands) are evident in our 3I reflectance spectrum at the current date.~However, we note a mild depression at 6000-7000\,\AA\ (see top right inset in Fig.~\ref{fig:refspec}) using the SOLSPEC normalization.~If true, this may indicate the onset of mineralogical features \citep[such as hydrated silicate bands; e.g.][]{Seccull24}, which would be consistent with 3I's location ($4.4$\,AU; see Fig.~\ref{fig:thevol}) and may represent material yet too subtle/buried under the dominant continuum.~We encourage further observations to corroborate or reject this tentative evidence.~In general, we observe that within the signal-to-noise range, the spectrum of 3I is smooth and devoid of obvious narrow absorptions, much like those of 1I at $r_h\!\simeq\!1.39$\,AU \citep{Ye17} and 2I at $r_h\!\simeq\!2.6$\,AU \citep{deLeon20}.

We observe no evidence of major gas emission from CN, C$_3$, C$_2$, CO$^+$, and \mbox{[O\,\textsc{i}]}, consistent with the non-detection of volatiles in Solar System comets at similar heliocentric distances and thermal conditions (Fig.~\ref{fig:thevol}).~In contrast, at 3.8\,AU, thermal models predict ice sublimation, as seen in Comet Hale-Bopp \citep[Fig.~\ref{fig:refspec}, bottom left zoom-in;][]{Rauer03}.~We also find no evidence for emission at 3840-80\,\AA\ in the 1-FWHM and 2-FWHM extracted spectrum (Fig.~\ref{fig:refspec}, bottom right zoom-in), indicating that there is currently no presence of CN (cyanogen) in the coma.~A broad absorption feature at 6000-7000\,\AA\ appears weakly in the 2-FWHM extraction using the SOLSPEC normalization (Fig.~\ref{fig:refspec}, upper right zoom-in), and may suggest the presence of NH$_2$ (amidogen) beyond the nucleus.~However, deeper spectra are needed to corroborate these results.~These weak or non-detections of volatiles imply that, at the time of observations, only very low levels of gas (primarily NH$_3$ and CO-bearing compounds) are being released, perhaps not surprising given that 3I was at $4.4$\,AU.~Even so, the reflectance continuum and a growing coma \citep{Chandler25, Opitom25, Seligman25} provide critical clues to its surface composition. 

\section{Discussion} 

\subsection{Thermal‐Evolution Model for 3I/ATLAS}
\label{ln:thevol}
We performed thermal evolution modeling of 3I using a numerical conduction model, similar to the method outlined by \citet{Fitzsimmons18} in their analysis of 1I.~Our model computes instantaneous radiative-equilibrium surface temperatures assuming a Bond albedo of 0.04, a bolometric emissivity of 0.95, and a heliocentric distance-dependent solar flux.~Ephemerides and orbital elements necessary for the heliocentric distance and thermodynamics computations were retrieved from the JPL Horizons system and the Minor Planet Center online services \citep{Giorgini96}.~To capture the subsurface thermal response, we employ a 1D thermal conduction approximation, treating subsurface heat propagation as a diffusion-like process characterized by a thermal diffusivity typical for cometary regolith \citep{Groussin13, Fitzsimmons18}.~We note that the model self-consistently treats 3I's thermal evolution by explicitly incorporating its high velocity through the Solar System.~That said, 3I's exact composition, coma and nucleus morphology, rotation, and thermal diffusion variance in surface and subsurface materials may alter these predictions.~We, therefore, defer a more detailed study to future work.

\begin{figure}[t!]
\centering
\includegraphics[width=1\columnwidth]{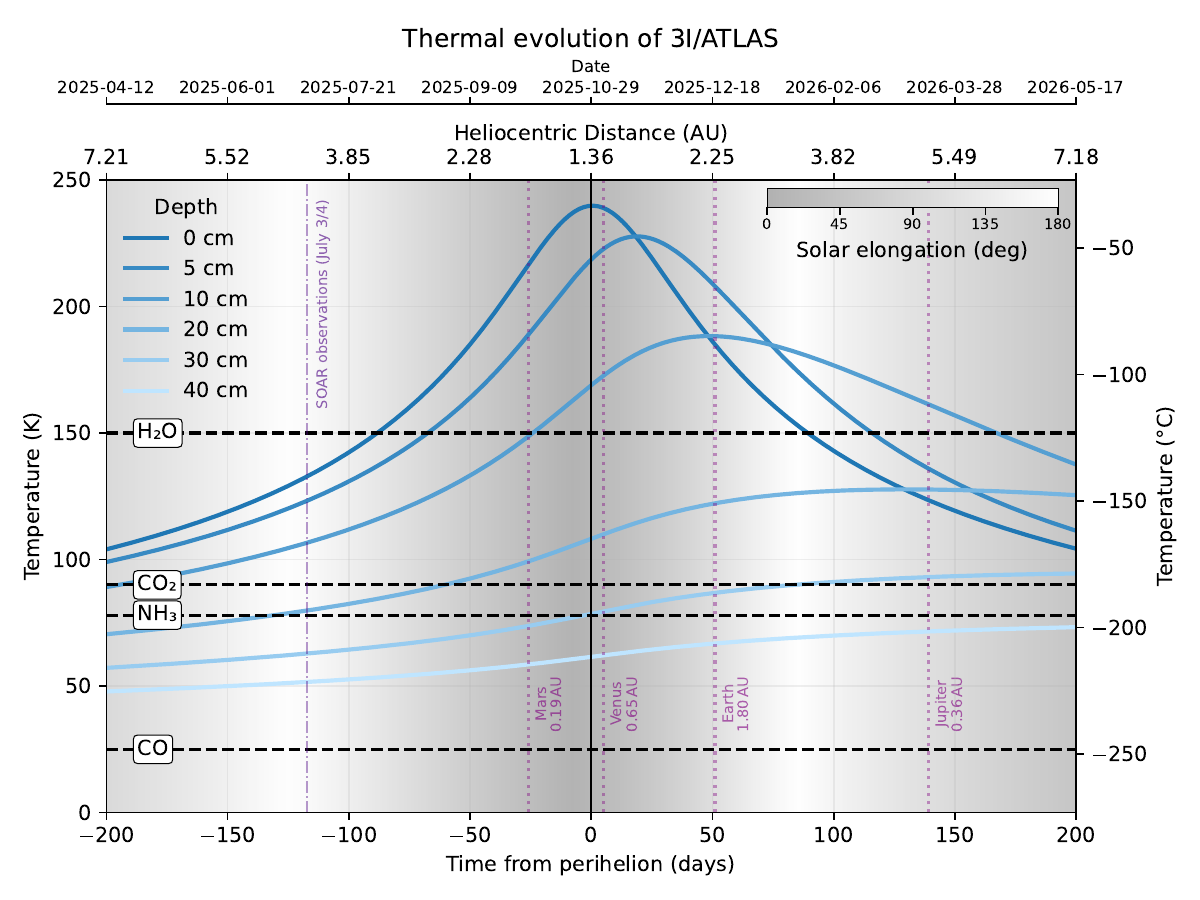}
\caption{Modeled thermal evolution of 3I during its perihelion passage.~Solid curves show the time–temperature profiles at multiple depths (0–40 cm; see legend) obtained with the 1D conduction model described in Section \ref{ln:thevol}. All subsurface layers were initialized at $T_{\rm ISM}\!=\!10$ K at $t=-10^{5}$ d and pre-conducted to the instantaneous radiative-equilibrium surface temperature prior to the plotted window. The solid vertical line marks the epoch of perihelion ($T_{\rm p}=2025$ Oct 29, see top axis). Dashed horizontal lines indicate vacuum sublimation thresholds for selected volatiles (CO, NH$_3$, CO$_2$, and H$_2$O). A twin right-hand axis gives the corresponding temperature in $^\circ$C.~The bottom x-axis shows the temporal perihelion offset in days, while the upper x-axes provide the corresponding heliocentric distance and date.~The gray shading encodes 3I's solar elongation, and the vertical dotted lines indicate encounters with major Solar system planets with their minimum passage distances shown in the labels.~The vertical dot-dashed line marks the date our SOAR observations were taken.
\label{fig:thevol}}
\end{figure}

Starting from an initial uniform internal temperature representative of the interstellar medium ($T_\mathrm{ISM}\!\approx\!10\,\mathrm{K}$) at $10^5$ days prior to perihelion, our model pre-conducts heat inward from the evolving surface boundary condition defined by radiative equilibrium.~By the time the simulation reaches the harsh inner Solar System conditions (typically $\pm200$ days around perihelion), the subsurface layers have thermally equilibrated to realistic internal temperature gradients. Sublimation thresholds for typical cometary volatiles (e.g.~CO, NH$_3$, CO$_2$, and H$_2$O; \citealt{Meech04}) were included to visually illustrate the potential activation and depletion depths of these species throughout 3I's perihelion passage.

Our model predicts that water ice remains stable at depths $\gtrsim\!15\!-\!20$\,cm.~H$_2$O, CO$_2$, NH$_3$ and CO could sublimate from shallower layers as 3I approaches perihelion, broadly consistent with the activity levels inferred from our early spectroscopy, showing very weak or no volatile emissions at 4.4\,AU.~The thermal model facilitates planing of future spectroscopic observations and direct comparison of 3I's predicted volatile activity, which is likely to pick up in August to October.

\subsection{Color Evolution}
\citet{Seligman25} reports on pre-discovery ZTF \citep{Bellm19} color evolution from $(g'\!-\!r')\!=\!0.42$ on 2025 May 22 (at phase angle $\alpha\!\simeq\!7^\circ$) to $(g'\!-\!r')\!=\!0.44$, close to the solar value, on 2025 June 18 (at phase angle $\alpha\simeq2.4^\circ$).~On our flux-calibrated spectra (taken at phase angle $\alpha\!\simeq\!2.9^\circ$) we measure 3I's $(g'\!-\!r')$ color in various filter systems, and obtain a SDSS color $(g'\!-\!r')\!=\!0.86\pm0.05$\,AB mag and for Pan-STARRS $(g-r)=0.73\pm0.05$\,AB mag. We also compute the corresponding LSST color $(g-r)=0.80\pm0.05$\,AB mag, which shows clear reddening.

Although the pre-discovery ZTF color evolution between May and June could be interpreted as evidence of a nascent coma, the opposition effect at such low phase angles could also partially or entirely explain it \citep[see e.g.][]{Rosenbush09}.~By the time our spectroscopic observations were taken, the evidence of reddening and a coma was already convincing \citep[e.g.][]{Opitom25}.~The similarities with 2I, whose dust color varied from moderately red to nearly neutral, depending on grain size and activity level during its approach \citep{Opitom19, Bolin20, Deam25}, suggests fresh dust from 3I’s coma having significantly red reflectance, whereas the underlying nucleus (or large-grain dust at large distance) may have a more neutral coloration.

\section{Conclusions}
Our early SOAR/Goodman HTS observations indicate that 3I/ATLAS shares the broadly reddish reflectance spectra of its interstellar predecessors, while potentially pushing to even redder slopes at blue wavelengths.~Its continuum slope variations \citep[see also][]{Yang25} and a weak absorption feature around 6000-7000\,\AA\  are reminiscent of the most primitive cometary and asteroidal surfaces \citep{Fraser22, Jewitt23}.~Similar to other spectroscopic observations at this epoch \citep[e.g.][]{Opitom25}, we find no indications for ice sublimation, consistent with 3I's thermal model state and the slow onset of subsurface material outgassing (Fig.~\ref{fig:thevol}).~The significant coma in combination with the lack of volatiles suggests that mechanisms of surface processing other than sublimation might be at play as 3I moves deeper into the Solar bubble \citep[e.g.][]{Reames99}. A particularly good candidate is solar-wind sputtering and UV desorption \citep[observed on 67P/Churyumov-Gerasimenko at $\sim$3\,AU;][]{Wurz15}, which may be releasing a significant rate of refractory materials at low speeds towards the Sun.~Indeed, \citet{Yang25} reported the presence of a 2$\mu$m water ice coma without an accompanying 1.5$\mu$m water ice band, favoring an early onset coma formation by solar wind sputtering. Additional mechanisms include  radioactive decay \citep{Prialnik90}, crystallization of amorphous ices \citep[e.g.][]{Prialnik92}, electrostatic dust lofting \citep{Wang16}, and photocatalytic exothermic reactions of the surface chemistry \citep{Cazaux16}, associated with an extended and cold ISM history \citep{Johnson87, Draine03}, all potentially contributing to the observed coma without measurable volatile activity.

One additional aspect in 3I's mildly paradoxical behavior, i.e.~growing a coma without detectable gas emission, involves progressive crust formation through multiple past stellar system encounters.~Although the latter is unlikely, given its expected dynamical history \citep[e.g.][]{Forbes25, Hopkins25}, we speculate that ISOs may experience multiple stellar system encounters early in their evolution, given that most stars form in a clustered fashion in so-called embedded clusters \citep{Lada03, MacLow04, PortegiesZwart10}.

~Laboratory studies demonstrate that refractory organic compounds accumulate as crusts during thermal processing \citep{Johnson87, Briani13}, while even thin dust mantles can reduce gas sublimation rates by factors of 5-50 \citep{Prialnik88}.~If 3I has undergone repeated stellar encounters over cosmic timescales, accumulated refractory crusts could suppress volatile emission while allowing continued dust ejection, explaining its unique activity profile among interstellar visitors.

As 3I is the first ISO estimated to be retained on Gyr timescales in the cold ISM \citep[see][]{Hopkins25}, and certainly the oldest cometary visitor yet observed, these crustal and dust‑ejection mechanisms are especially compelling.~The chemical nature of its surface crust may also offer insights into mechanisms posited for 1I/‘Oumuamua’s anomalous acceleration, e.g.~radiolytically‐produced H$_2$ in water ice \citep{Bergner23} and/or N$_2$ ice fragments \citep{Desch21}, thus, providing a pathway to compare interstellar visitors at different evolutionary states.

These characteristics underscore the scientific significance of 3I/ATLAS: by comparing its spectral properties with those of 1I/'Oumuamua, 2I/Borisov, and analogous Solar System small bodies (active comets, dormant comet nuclei, asteroids, TNOs, KBOs, etc.), we can begin to discern which traits are universal for planetesimals formed in other stellar nurseries. Dynamical models have suggested that 3I may originate from an old, thick-disk stellar population in the Milky Way \citep{Hopkins25}. If true, its red and refractory-rich surface may reflect eons of exposure in interstellar space \citep{Ferriere01, Draine03, Tielens08, Herbst09}.~Continued photometric monitoring and spectroscopic follow-up, especially as 3I approaches perihelion and its activity intensifies, will help us test formation hypotheses and sharpen our diagnostic tools as we gear up for the LSST era.

\begin{acknowledgments}
We are grateful to Wesley Fraser, Bin Yang, Michaël Marsset, and Michele Bannister for valuable discussions, and we thank the anonymous referee for their constructive comments, which improved the clarity of the manuscript.~We gratefully acknowledge support from the National Agency for Research and Development (ANID) grants:~CATA-Basal FB210003; Beca de Doctorado Nacional (RR, JPC).~This research has made use of data and/or services provided by the International Astronomical Union's Minor Planet Center, the National Aeronautics and Space Administration (NASA) Jet Propulsion Laboratory’s Horizons System, available at \url{https://ssd.jpl.nasa.gov/horizons/}, and the NASA/IPAC Extragalactic Database, which is funded by the NASA and operated by the California Institute of Technology.
\end{acknowledgments}

\facilities{SOAR (GoodmanHTS)}

\software{astropy \citep{astropy2013,astropy2018,astropy2022}, matplotlib \citep{Hunter07} }

\bibliography{3IATLAS}{}
\bibliographystyle{aasjournal}

\end{document}